\documentclass{Interspeech}

\interspeechcameraready 
\title{Voice Quality Dimensions as Interpretable Primitives for Speaking Style for Atypical Speech and Affect}

\author[affiliation={1}]{Jaya}{Narain}
\author[affiliation={1}]{Vasudha}{Kowtha}
\author[affiliation={1}]{Colin}{Lea}
\author[affiliation={1}]{Lauren}{Tooley}
\author[affiliation={1}]{Dianna}{Yee}
\author[affiliation={1}]{Vikramjit}{Mitra}
\author[affiliation={1}]{Zifang}{Huang}
\author[affiliation={1}]{Miquel}{Espi Marques}
\author[affiliation={1}]{Jon}{Huang}
\author[affiliation={1}]{Carlos}{Avendano}
\author[affiliation={1}]{Shirley}{Ren}

\affiliation{}{Apple}{USA}
\email{jnarain@apple.com}
\keywords{speech recognition, human-computer interaction, computational paralinguistics}

\usepackage{comment}
\usepackage{dblfloatfix}
\usepackage{cite}

\begin{document}

\maketitle

\begin{abstract}
Perceptual voice quality dimensions describe key characteristics of atypical speech and other speech modulations. Here we develop and evaluate voice quality models for seven voice and speech dimensions (intelligibility, imprecise consonants, harsh voice, naturalness, monoloudness, monopitch, and breathiness). Probes were trained on the public Speech Accessibility (SAP) project dataset with 11,184 samples from 434 speakers, using embeddings from frozen pre-trained models as features. We found that our probes had both strong performance and strong generalization across speech elicitation categories in the SAP dataset. We further validated zero-shot performance on additional datasets, encompassing unseen languages and tasks: Italian atypical speech, English atypical speech, and affective speech. The strong zero-shot performance and the interpretability of results across an array of evaluations suggests the utility of using voice quality dimensions in speaking style-related tasks.
\end{abstract}

\section{Introduction and Related Work}

Voice quality dimensions (VQDs) capture information about a speaker's style and voice related to anatomy, vocal tract configuration, and learned components \cite{vqdef}.  Voice quality can be impacted by disorders directly related to speech including structural (e.g., voice nodules) and neurogenic conditions (e.g., spasmodic dysphonia), as well as by other long- and short-term health conditions (e.g., neuromotor conditions like ALS, and Parkinson’s and transient illnesses like colds and sore throats) \cite{green2024artificial, bot2016mpower}.  Day-to-day voice fatigue and overuse also impacts speech, causing both short- and long-term changes \cite{wingate2007treatment}.  

Evaluations by speech-language pathologists (SLP) include multiple dimensions of vocal attributes as in the CAPE-V \cite{kempster2009consensus} and GRBAS \cite{aud_perc} which rate severity, roughness, breathiness, asthenia, strain, pitch, and loudness.  Modeling approaches that use an array of perceptual voice quality dimensions (VQDs) could provide a holistic view of voice for a variety of applications; however, this area has been under-explored.  For instance, automatic characterization of VQDs could enable improved data curation for and evaluation of automatic speech recognition (ASR) and could also help identify and characterize speech differences - for instance related to wellbeing (e.g., respiratory conditions, fatigue, and affect).  

Prior work in speech science has studied voice quality descriptors \cite{kreiman2024information}, including as they relate to affect \cite{becker2022beyond, gobl2003role}, atypical speech \cite{bele2007dimensionality}, speaker ID \cite{baumann2010perceptual}, and clinical uses \cite{roy2013evidence}.  In the computational and machine learning domain, there has been limited work on models directly targeting interpretable VQDs.  A study by \cite{xie2018machine} showed that models could predict GRBAS \cite{aud_perc} scores on a dataset of n=20 speakers, though it did not evaluate generalization or use public datasets.  Other work has largely focused on acoustic feature correlates to voice properties, as in \cite{dhamyal2024objective, eskenazi1990acoustic, shue2010voice} and on single dimensions of atypical speech, like dysfluency \cite{romana2022enabling}, paraphasia \cite{perez2023seq2seq}, and intelligibility \cite{venugopalan2023speech}.
Prior work has extensively explored models and features for task-specific speaking styles -- for example health (e.g., mental health \cite{ringeval2019avec}, fatigue \cite{gao2022rapid}, respiratory conditions \cite{talkar2023dissociating}, and ALS and other neuromotor and neurologic conditions \cite{vieira2022machine, luz2021alzheimer}), and affect \cite{ringeval2019avec, mitra2024investigating, goncalves2024odyssey}.  These works generally model a single factor, and often use a particular type of data and/or task (e.g., English voice commands).  Modeling approaches included using embeddings as features \cite{venugopalan2023speech, mitra2024investigating}, fine-tuning \cite{romana2022enabling, goncalves2024odyssey}, and supervised models including classical algorithms \cite{ringeval2019avec, gao2022rapid, talkar2023dissociating}, convolutional models \cite{vieira2022machine}, and transformers \cite{perez2023seq2seq}.  In the realm of general non-semantic speech understanding, prior work
\cite{shor2022universal, lin2023utility} has proposed benchmarks and models, with a focus on fine-tuning for specific tasks but not on zero-shot transfer or interpretable VQD primitives.    

Here we focus on seven VQDs (intelligibility, imprecise consonants, harsh voice, naturalness, monoloudness, monopitch, and breathiness) as interpretable primitives, as learned from atypical speech data which includes high expressivity across dimensions.  We investigate three primary research questions: (1) What is the relative performance of large pre-trained embedding models as feature extractors for predicting VQDs for atypical speech?  (2) Do models generalize (zero-shot) across speakers, datasets, tasks, and languages for atypical speech?  (3) Do models and dimensions generalize (zero-shot) beyond atypical speech to an affect dataset?  

Our models are trained on English data from the Speech Accessibility Project (SAP) dataset, and are evaluated on SAP data and on three out-of-domain datasets: English atypical speech dataset, an Italian atypical dataset, and an affect dataset -- covering different phrase contents, speaker distributions, languages, and tasks.  In the long run, methods that incorporate VQDs in modeling could improve accessibility and enable new speech technologies by helping isolate specific speaking styles, improving interpretability, and improving robustness by identifying aggressors and styles that might be easily confused with each other -- for instance, some VQDs may be more related to atypical speech than affect and some may be impacted by both.

\section{Methods}

\subsection{Training data and labels}
Probes were trained using the Speech Accessibility Project \cite{hasegawa2024community} a publicly available dataset of atypical speech.  We used the subset of data with annotations from speech-language pathologists, which included 11,184 samples from 434 speakers (n=284 with Parkinson's Disease, 78 with Cerebral Palsy, 53 with ALS, 16 with Down Syndrome, 2 with Ataxic Dysarthria, 1 with Flaccid Dyarthria) from three speech categories (see \cite{hasegawa2024community} for examples): digital voice commands (\textit{n}=3,797; avg. length 5.00s, std. 4.65s), novel sentences (\textit{n}=3,838; avg. length 9.03s, std. 4.42s); and spontaneous speech samples (\textit{n}=3,549; avg. length 25.94s, std. 21.47s).  We used the dataset's speaker-stratified train (\textit{n}=8,016), validation (\textit{n}=1,194), and test splits (\textit{n}=2,086).  Each sample was annotated for each dimension from 1-7, with 1 corresponding to typical and 7 to strongly atypical speech in that dimension.  The dimensions used in this study were chosen because they had been rated for most samples, and because at least 10\% of samples had a rating $\geq$ 2 for each dimension.  

\begin{figure}[h!]
  \centering
  \includegraphics[width=0.6\linewidth]{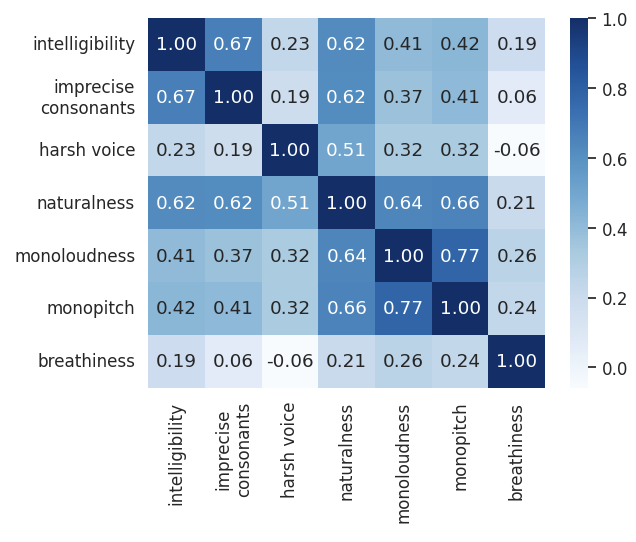}
  \caption{Pearson correlations between annotations}
  \label{fig:annotation_correlations}
\end{figure}

\begin{figure}[t]
\centering

  \includegraphics[width=1\linewidth]{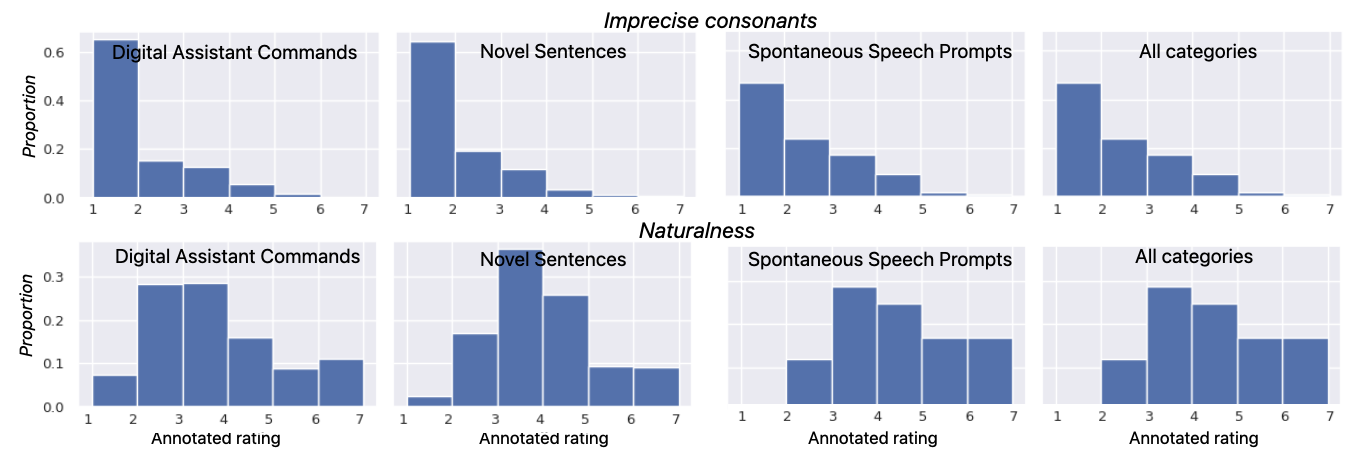}
  \caption{Distributions of annotations for imprecise consonants and naturalness for each speech category}
  \label{fig:sap_annotation_distributions.png}
\end{figure}

\begin{table*}[!b]
\centering
\caption{Spearman correlations and AUC per VQD (95\% CIs from bootstrapping). *SpICE was trained specifically for intelligibility}
\label{tab:spearman_all}
\begin{tabular}{lccccc|ccccc}
\hline
                                                        & \multicolumn{5}{c|}{\textbf{Spearman correlation}}                                                                                                                                                                                                                                                                                                     & \multicolumn{5}{c}{\textbf{AUC}}                                                                                                                                                                                                                                                                                                                       \\ \hline
                                                        & \multicolumn{1}{l}{CLAP}                                          & \multicolumn{1}{l}{HuBERT}                                        & \multicolumn{1}{l}{\begin{tabular}[c]{@{}l@{}}HuBERT\\ ASR\end{tabular}} & \multicolumn{1}{l}{\begin{tabular}[c]{@{}l@{}}Raw-\\ Net3\end{tabular}} & \multicolumn{1}{l|}{SpICE}                                & \multicolumn{1}{l}{CLAP}                                          & \multicolumn{1}{l}{HuBERT}                                        & \multicolumn{1}{l}{\begin{tabular}[c]{@{}l@{}}HuBERT\\ ASR\end{tabular}} & \multicolumn{1}{l}{\begin{tabular}[c]{@{}l@{}}Raw-\\ Net3\end{tabular}} & \multicolumn{1}{l}{SpICE}                                 \\ \hline
Intellig.                                               & \begin{tabular}[c]{@{}c@{}}.45\\ (.41, .48)\end{tabular}          & \begin{tabular}[c]{@{}c@{}}.49\\ (.49, .53)\end{tabular}          & \textbf{\begin{tabular}[c]{@{}c@{}}.52\\ (.48, .55)\end{tabular}}        & \begin{tabular}[c]{@{}c@{}}.33\\ (.30, .37)\end{tabular}                & \begin{tabular}[c]{@{}c@{}}.41*\\ (.37, .44)\end{tabular} & \begin{tabular}[c]{@{}c@{}}.78\\ (.76, .80)\end{tabular}          & \begin{tabular}[c]{@{}c@{}}.81\\ (.78, .83)\end{tabular}          & \textbf{\begin{tabular}[c]{@{}c@{}}.82\\ (.80, .84)\end{tabular}}        & \begin{tabular}[c]{@{}c@{}}.72\\ (.69, .74)\end{tabular}                & \begin{tabular}[c]{@{}c@{}}.75*\\ (.72, .77)\end{tabular} \\
\begin{tabular}[c]{@{}l@{}}Impr.\\ conson.\end{tabular} & \begin{tabular}[c]{@{}c@{}}.58\\ (.55, .61)\end{tabular}          & \begin{tabular}[c]{@{}c@{}}.68\\ (.66, .70)\end{tabular}          & \textbf{\begin{tabular}[c]{@{}c@{}}.69\\ (.67, .70)\end{tabular}}        & \begin{tabular}[c]{@{}c@{}}.51\\ (.47, .55)\end{tabular}                & \begin{tabular}[c]{@{}c@{}}.52\\ (.49, .54)\end{tabular}  & \begin{tabular}[c]{@{}c@{}}.84\\ (.82, .86)\end{tabular}          & \begin{tabular}[c]{@{}c@{}}.90\\ (.89, .91)\end{tabular}          & \textbf{\begin{tabular}[c]{@{}c@{}}.91\\ (.90, .92)\end{tabular}}        & \begin{tabular}[c]{@{}c@{}}.81\\ (.79, .83)\end{tabular}                & \begin{tabular}[c]{@{}c@{}}.82\\ (.80, .84)\end{tabular}  \\
\begin{tabular}[c]{@{}l@{}}Harsh\\ voice\end{tabular}   & \textbf{\begin{tabular}[c]{@{}c@{}}.63\\ (.59, .65)\end{tabular}} & \begin{tabular}[c]{@{}c@{}}.55\\ (.52, .58)\end{tabular}          & \begin{tabular}[c]{@{}c@{}}.48\\ (.48, .51)\end{tabular}                 & \begin{tabular}[c]{@{}c@{}}.55\\ (.51, .58)\end{tabular}                & \begin{tabular}[c]{@{}c@{}}.12\\ (.07, .16)\end{tabular}  & \textbf{\begin{tabular}[c]{@{}c@{}}.86\\ (.85, .89)\end{tabular}} & \begin{tabular}[c]{@{}c@{}}.83\\ (.81, .85)\end{tabular}          & \begin{tabular}[c]{@{}c@{}}.80\\ (.78, .82)\end{tabular}                 & \begin{tabular}[c]{@{}c@{}}.77\\ (.74, .79)\end{tabular}                & \begin{tabular}[c]{@{}c@{}}.53\\ (.49, .57)\end{tabular}  \\
\begin{tabular}[c]{@{}l@{}}Natural\\ -ness\end{tabular} & \begin{tabular}[c]{@{}c@{}}.66\\ (.63, .68)\end{tabular}          & \begin{tabular}[c]{@{}c@{}}.72\\ (.69, .74)\end{tabular}          & \begin{tabular}[c]{@{}c@{}}.70\\ (.67, .72)\end{tabular}                 & \begin{tabular}[c]{@{}c@{}}.51\\ (.47, .54)\end{tabular}                & \begin{tabular}[c]{@{}c@{}}.47\\ (.44, .50)\end{tabular}  & \begin{tabular}[c]{@{}c@{}}.86\\ (.84, .88)\end{tabular}          & \textbf{\begin{tabular}[c]{@{}c@{}}.89\\ (.87, .90)\end{tabular}} & \begin{tabular}[c]{@{}c@{}}.89\\ (.88, .91)\end{tabular}                 & \begin{tabular}[c]{@{}c@{}}.78\\ (.75, .80)\end{tabular}                & \begin{tabular}[c]{@{}c@{}}.81\\ (.78, .83)\end{tabular}  \\
\begin{tabular}[c]{@{}l@{}}Mono\\ loud.\end{tabular}    & \textbf{\begin{tabular}[c]{@{}c@{}}.62\\ (.59, .64)\end{tabular}} & \begin{tabular}[c]{@{}c@{}}.61\\ (.59, .53)\end{tabular}          & \begin{tabular}[c]{@{}c@{}}.56\\ (.54, .59)\end{tabular}                 & \begin{tabular}[c]{@{}c@{}}.41\\ (.38, .44)\end{tabular}                & \begin{tabular}[c]{@{}c@{}}.34\\ (.30, .38)\end{tabular}  & \textbf{\begin{tabular}[c]{@{}c@{}}.82\\ (.80, .84)\end{tabular}} & \begin{tabular}[c]{@{}c@{}}.81\\ (.79, .82)\end{tabular}          & \begin{tabular}[c]{@{}c@{}}.80\\ (.78, .82)\end{tabular}                 & \begin{tabular}[c]{@{}c@{}}.72\\ (.70, .74)\end{tabular}                & \begin{tabular}[c]{@{}c@{}}.68\\ (.65, .70)\end{tabular}  \\
\begin{tabular}[c]{@{}l@{}}Mono\\ -pitch\end{tabular}   & \begin{tabular}[c]{@{}c@{}}.56\\ (.52, .58)\end{tabular}          & \textbf{\begin{tabular}[c]{@{}c@{}}.57\\ (.54, .60)\end{tabular}} & \begin{tabular}[c]{@{}c@{}}.53\\ (.50, .56)\end{tabular}                 & \begin{tabular}[c]{@{}c@{}}.31\\ (.27, .34)\end{tabular}                & \begin{tabular}[c]{@{}c@{}}.30\\ (.24, .33)\end{tabular}  & \textbf{\begin{tabular}[c]{@{}c@{}}.82\\ (.81, .84)\end{tabular}} & \begin{tabular}[c]{@{}c@{}}.81\\ (.80, .84)\end{tabular}          & \begin{tabular}[c]{@{}c@{}}.81\\ (.79, .83)\end{tabular}                 & \begin{tabular}[c]{@{}c@{}}.70\\ (.67, .82)\end{tabular}                & \begin{tabular}[c]{@{}c@{}}.69\\ (.67, .71)\end{tabular}  \\
\begin{tabular}[c]{@{}l@{}}Breath\\ -iness\end{tabular} & \begin{tabular}[c]{@{}c@{}}.31\\ (.39, .35)\end{tabular}          & \textbf{\begin{tabular}[c]{@{}c@{}}.35\\ (.31, .38)\end{tabular}} & \begin{tabular}[c]{@{}c@{}}.30\\ (.26, .33)\end{tabular}                 & \begin{tabular}[c]{@{}c@{}}.23\\ (.20, .27)\end{tabular}                & \begin{tabular}[c]{@{}c@{}}.21\\ (.17, .25)\end{tabular}  & \begin{tabular}[c]{@{}c@{}}.82\\ (.79, .84)\end{tabular}          & \textbf{\begin{tabular}[c]{@{}c@{}}.82\\ (.80, .85)\end{tabular}} & \begin{tabular}[c]{@{}c@{}}.81\\ (.78, .83)\end{tabular}                 & \begin{tabular}[c]{@{}c@{}}.74\\ (.70, .77)\end{tabular}                & \begin{tabular}[c]{@{}c@{}}.68\\ (.65, .71)\end{tabular}  \\ \hline
\textit{Avg./std.}                                      & \multicolumn{1}{l}{.54/.11}                                       & \multicolumn{1}{l}{.57/.11}                                       & \multicolumn{1}{l}{.54/.13}                                              & \multicolumn{1}{l}{.41/.11}                                             & \multicolumn{1}{l|}{.33/.13}                              & \multicolumn{1}{l}{0.83/.03}                                      & \multicolumn{1}{l}{0.84/.04}                                      & \multicolumn{1}{l}{0.84/.04}                                             & \multicolumn{1}{l}{0.75/.04}                                            & \multicolumn{1}{l}{0.71/.09}                             
\end{tabular}
\end{table*}
Fig. \ref{fig:annotation_correlations} shows correlations between ratings in the dataset.  The annotated dimensions are not perfectly correlated -- each captures some unique information, suggesting the utility of considering each dimension as a modeling target. Some have moderately strong correlations (e.g., monoloudness \& monopitch), while others have weak to no correlation (e.g., harsh voice \& breathiness).  Annotation distributions vary by the dimension itself and based on speech category type (Fig. \ref{fig:sap_annotation_distributions.png}), which motivates assessing generalizability in models between speech categories, as does prior work in \cite{tobin2024automatic}.

\subsection{Embedding extraction}

We used frozen audio embedding models as feature extractors:

\begin{itemize}
    \item HuBERT Large (315M parameters; emb. dim 1024), HuBERT model \cite{hsu2021hubert}, self-supervised model trained via masking
    \item HuBERT Large ASR (315M parameters; emb. dim 1024), HuBERT model fine-tuned for ASR \cite{hsu2021hubert}
    \item Internal CLAP model (86M parameters; emb. dim 784), internal audio embedding model trained with masking followed by language alignment on audio samples with human-annotated captions describing sound
    \item RawNet3 from ESPNet-SPK (28M parameters; emb. dim 192) \cite{jung2024espnet}, pre-trained speaker identification model

\end{itemize}

\begin{figure}[t]
  \centering
  \includegraphics[width=0.8\linewidth]{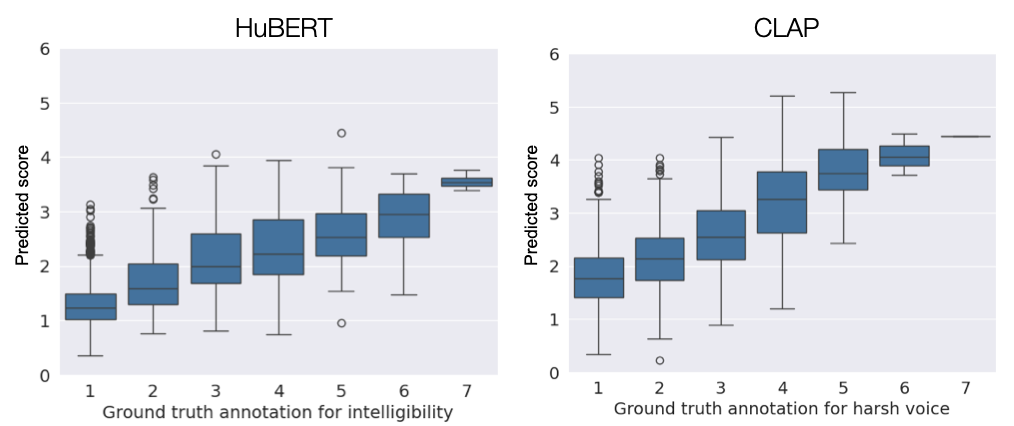}
  \caption{Regression probes, showing clear progression of predicted scores with rated severity}

  \label{fig:regression_output_plots}
\end{figure}

HuBERT Large and HuBERT Large ASR were included to investigate the impact of ASR-targeted fine-tuning and to compare with the CLAP training strategy. RawNet3 was included despite its smaller size to evaluate the use of a model trained for speaker ID, which could relate to VQDs as they may be tied to speaker characteristics.  Leading and trailing silence was trimmed prior to embedding extraction for all samples a wav2vec2 forced alignment model \cite{wav2vecfa} and annotated transcriptions (alignments were validated by spot checking a subset of data).  We trained Lasso models as regression probes and logistic regression models as binary classification probes for each VQD, using the validation set for selecting the regularization parameter value.  We converted the annotated scores to a binary label for classification, aiming for near 20\% positive labels (selected to stabilize task difficulty and enable better comparisons between model performance across dimensions). We also trained NN probes, but found that linear probes had consistently better performance.  

To benchmark with prior work, we also evaluated the pre-trained SpICE model \cite{venugopalan2023speech}, trained on 550K+ utterances to classify speech intelligibility using a wav2vec2 backbone.  We converted the multi-class prediction from the probe to a score by computing a label-weighted average of the model output.

\begin{table*}[!h]
\centering
\caption{Regression Spearman correlations avgs.(stds.) across VQDs, by training and evaluation speech category.   
}
\label{tab:speech_category_generalization}
\begin{tabular}{l|lll|lll|lll|lll}
\begin{tabular}[c]{@{}l@{}}Eval.\\ set\end{tabular} & \textbf{DAC}                                                 & \textbf{\begin{tabular}[c]{@{}l@{}}Novel\\ Sent.\end{tabular}} & \textbf{\begin{tabular}[c]{@{}l@{}}Spont.\\ Speech\end{tabular}} & \textbf{DAC}                                                 & \textbf{\begin{tabular}[c]{@{}l@{}}Novel\\ Sent.\end{tabular}} & \textbf{\begin{tabular}[c]{@{}l@{}}Spont.\\ Speech\end{tabular}} & \textbf{DAC}                                                  & \textbf{\begin{tabular}[c]{@{}l@{}}Novel\\ Sent.\end{tabular}} & \textbf{\begin{tabular}[c]{@{}l@{}}Spont.\\ Speech\end{tabular}} & \textbf{DAC}                                                 & \textbf{\begin{tabular}[c]{@{}l@{}}Novel\\ Sent.\end{tabular}} & \textbf{\begin{tabular}[c]{@{}l@{}}Spont.\\ Speech\end{tabular}} \\ \hline
\textit{}                                           & \multicolumn{3}{l|}{\textit{Trained on all data}}                                                                                                                                                & \multicolumn{3}{l|}{\textit{Trained on DAC}}                                                                                                                                                     & \multicolumn{3}{l|}{\textit{Trained on novel sent.}}                                                                                                                                              & \multicolumn{3}{l}{\textit{Trained on spont. speech}}                                                                                                                                            \\ \hline
HuBERT                                              & \begin{tabular}[c]{@{}l@{}}.51\\ (.12)\end{tabular}          & \textbf{\begin{tabular}[c]{@{}l@{}}.54\\ (.12)\end{tabular}}   & \textbf{\begin{tabular}[c]{@{}l@{}}.60\\ (.13)\end{tabular}}     & \begin{tabular}[c]{@{}l@{}}.47\\ (.11)\end{tabular}          & \begin{tabular}[c]{@{}l@{}}.42\\ (.10)\end{tabular}            & \textbf{\begin{tabular}[c]{@{}l@{}}.52\\ (.10)\end{tabular}}     & \begin{tabular}[c]{@{}l@{}}.29\\ (.09)\end{tabular}           & \textbf{\begin{tabular}[c]{@{}l@{}}.53\\ (.12)\end{tabular}}   & \textbf{\begin{tabular}[c]{@{}l@{}}.53\\ (.15)\end{tabular}}     & \begin{tabular}[c]{@{}l@{}}.42\\ (.10)\end{tabular}          & \begin{tabular}[c]{@{}l@{}}.45\\ (.13)\end{tabular}            & \textbf{\begin{tabular}[c]{@{}l@{}}.57\\ (.13)\end{tabular}}     \\
HuB-ASR                                             & \textbf{\begin{tabular}[c]{@{}l@{}}.52\\ (.15)\end{tabular}} & \begin{tabular}[c]{@{}l@{}}.50 \\ (.11)\end{tabular}           & \begin{tabular}[c]{@{}l@{}}.55\\ (.14)\end{tabular}              & \textbf{\begin{tabular}[c]{@{}l@{}}.49\\ (.15)\end{tabular}} & \begin{tabular}[c]{@{}l@{}}.40\\ (.12)\end{tabular}            & \begin{tabular}[c]{@{}l@{}}.49\\ (.13)\end{tabular}              & \begin{tabular}[c]{@{}l@{}}.34\\ (.14)\end{tabular}           & \begin{tabular}[c]{@{}l@{}}.51\\ (.13)\end{tabular}            & \begin{tabular}[c]{@{}l@{}}.47\\ (.12)\end{tabular}              & \textbf{\begin{tabular}[c]{@{}l@{}}.43\\ (.16)\end{tabular}} & \begin{tabular}[c]{@{}l@{}}.41\\ (.13)\end{tabular}            & \begin{tabular}[c]{@{}l@{}}.54\\ (.15)\end{tabular}              \\
CLAP                                                & \begin{tabular}[c]{@{}l@{}}.49\\ (.13)\end{tabular}          & \begin{tabular}[c]{@{}l@{}}.53\\ (.12)\end{tabular}            & \begin{tabular}[c]{@{}l@{}}.55\\ (.12)\end{tabular}              & \begin{tabular}[c]{@{}l@{}}.49\\ (.12)\end{tabular}          & \textbf{\begin{tabular}[c]{@{}l@{}}.45\\ (.13)\end{tabular}}   & \begin{tabular}[c]{@{}l@{}}.50\\ (.11)\end{tabular}              & \begin{tabular}[c]{@{}l@{}}.31\\ (.15)\end{tabular}           & \begin{tabular}[c]{@{}l@{}}.50\\ (.13)\end{tabular}            & \begin{tabular}[c]{@{}l@{}}.49\\ (.13)\end{tabular}              & \begin{tabular}[c]{@{}l@{}}.42\\ (.11)\end{tabular}          & \textbf{\begin{tabular}[c]{@{}l@{}}.49\\ (.09)\end{tabular}}   & \begin{tabular}[c]{@{}l@{}}.53\\ (.09)\end{tabular}              \\
RawNet3                                             & \begin{tabular}[c]{@{}l@{}}.40\\ (.11)\end{tabular}          & \begin{tabular}[c]{@{}l@{}}.39\\ (.14)\end{tabular}            & \begin{tabular}[c]{@{}l@{}}.42\\ (.12)\end{tabular}              & \begin{tabular}[c]{@{}l@{}}.35\\ (.12)\end{tabular}          & \begin{tabular}[c]{@{}l@{}}.27\\ (.17)\end{tabular}            & \begin{tabular}[c]{@{}l@{}}.37\\ (.14)\end{tabular}              & \textbf{\begin{tabular}[c]{@{}l@{}}.35 \\ (.09)\end{tabular}} & \begin{tabular}[c]{@{}l@{}}.36\\ (.12)\end{tabular}            & \begin{tabular}[c]{@{}l@{}}.38\\ (.11)\end{tabular}              & \begin{tabular}[c]{@{}l@{}}.36\\ (.11)\end{tabular}          & \begin{tabular}[c]{@{}l@{}}.34\\ (.15)\end{tabular}            & \begin{tabular}[c]{@{}l@{}}.41\\ (.13)\end{tabular}             
\end{tabular}
\end{table*}
\begin{figure}[h!]
  \centering
  \includegraphics[width=\linewidth]{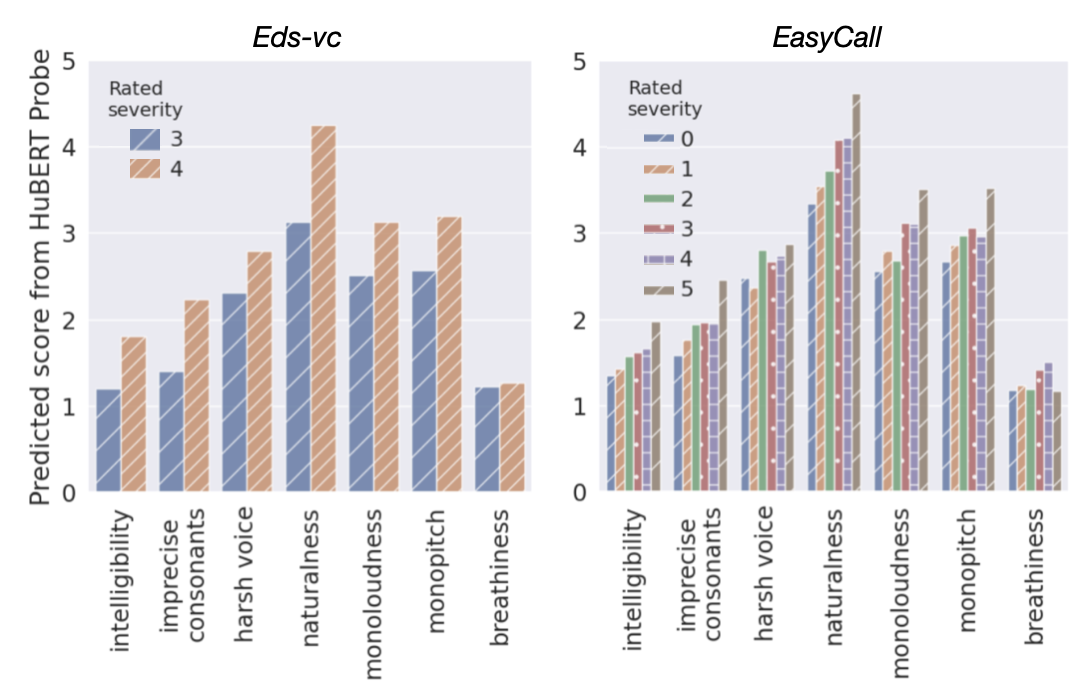}
  \caption{Zero-shot predictions from the HuBERT probe for each voice quality dimension on the Eds-vc dataset and on the EasyCall dataset, stratified by rated speech severity}
  \label{fig:external_severity}
\end{figure}

\subsection{Generalization}

We investigated generalization between speech categories (by training probes independently on each speech category), and generalization to new datasets - including to a language not seen during probe training.  To evaluate generalization to new datasets, probes were used for zero-shot predictions on two datasets of atypical speech: Eds-vc, an internal \textit{E}nglish \textit{d}ysarthric \textit{s}peech dataset of \textit{v}oice \textit{c}ommands, and the EasyCall Italian dataset \cite{turrisi2021easycall}.
Eds-vc had 35,661 samples from \textit{n}=32 speakers with moderate to severe dysarthric speech (approximately 75\% speakers with CP and the remaining with unknown etiologies).  Each speaker contributed ten repetitions of 105 phrases spread across five recording sessions, and speaker severity was annotated by an SLP.  EasyCall \cite{turrisi2021easycall} has 21,361 Italian phrases recorded from \textit{n}=51 participants (\textit{n}=31 with dysarthria).  Each speaker contributed 20 phrases and 46 words across 2-8 recording sessions.  Severity labels are from assessments by a neurologist using the Therapy Outcome Measure.  Note that compared to the data used to train each probe, the Eds-vc and EasyCall datasets had different participant etiology distributions, recording protocols, and content and that the EasyCall dataset was in a different language.

\subsection{Zero shot affect exploration}

To explore the broader utility of the VQD primitives, we analyzed zero-shot predictions of the modeled VQD on an affective speech using 'high' intensity samples from the Ryerson Audio-Visual Database of Emotional Speech and Song (RAVDESS) \cite{livingstone2018ryerson} dataset.  RAVDESS is an acted affect dataset with seven categorical emotions: calm, happy, sad, angry, fearful, disgust, surprised.  We used probes trained only on the SAP dataset (which has no elicited affective content) to generate predictions for each VQD.  We tabulated the average dimensional score for each categorical emotion. 

\begin{table}[!b]
\setlength{\tabcolsep}{2.5pt}
\caption{Zero-shot AUC for binary classification of atypical speech severity on out-of-domain datasets, per sample.  The 95\% CI was within 0.01 of the reported score for all results.}
\label{tab:auc_external}
\begin{tabular}{llll|lll}
                & HuB.         & CLAP         & SpICE        & HuB.          & CLAP          & SpICE \\ \hline
                & \multicolumn{3}{l|}{\textit{Eds-vc}}       & \multicolumn{3}{l}{\textit{EasyCall}} \\ \hline
Sum (all dims.) & \textbf{.89} & .87          & .79          & .72           & \textbf{.78}  & .66   \\ \hline
Intelligibility & .85          & \textbf{.88} & .79          & .69           & \textbf{.70}  & .58   \\
Impr. cons.     & .85          & \textbf{.87} & .78          & .66           & \textbf{.71}  & .67   \\
Harsh voice     & .69          & .73          & \textbf{.79} & \textbf{.67}  & .65           & .66   \\
Naturalness     & \textbf{.88} & .83          & .79          & .70           & \textbf{.74}  & .65   \\
Monoloud.       & \textbf{.84} & .80          & .64          & .65           & \textbf{.74}  & .66   \\
Monopitch       & \textbf{.83} & .77          & .79          & .59           & \textbf{.70}  & .43   \\
Breathiness     & \textbf{.59} & .57          & .31          & .67           & \textbf{.66}  & .66   \\ \hline
\end{tabular}
\end{table}

\begin{figure}[h!]
  \centering
  \includegraphics[width=\linewidth]{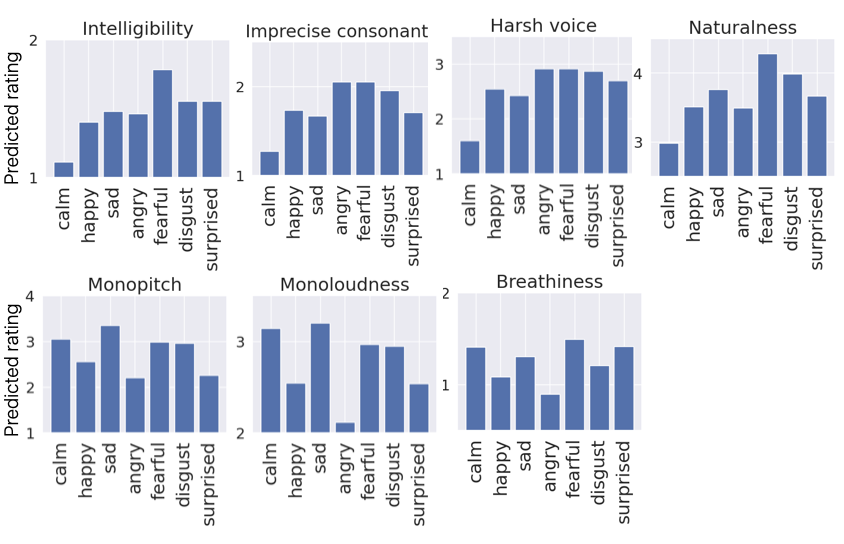}
  \caption{Zero-shot predictions for each voice quality dimension on RAVDESS using probes trained on HuBERT for each categorical emotion}
  \label{fig:hubert_ravdess_all}
\end{figure}

\ 


\section{Results and Discussion}

\hspace{\parindent}\textbf{Embeddings and Speech Category Generalization.} Table \ref{tab:spearman_all} lists Spearman correlations (similar trends were observed for the R2 and mean average error) between the regression probe prediction and ground truth annotation and the AUC for each classification probe, trained and evaluated on all speech categories.  Figure \ref{fig:regression_output_plots} shows regression results for two VQDs.  Table \ref{tab:speech_category_generalization} has the average Spearman correlation across VQDs by speech category. HuBERT ASR features had strongest performance for dimensions tied to pronunciation, like intelligibility and imprecise consonants (Table \ref{tab:spearman_all}), which may be because ASR is strongly tied to pronunciations.  CLAP had strong performance for harshness, which is related strongly to tone.  Probes generalized to unseen speech category types across all models (though performance was better when all categories were in training), showing utility of the probes on out of distribution data (Table \ref{tab:speech_category_generalization}).  Interestingly, probes trained only on novel sentences had poor generalization despite the training data having diverse phonetics.  This may be because reading novel sentences may elicit less representative speech than other categories which are more related to day-to-day speech.  Model performance was highest for spontaneous speech, even when it was not in training, perhaps due to its relatively longer sample lengths and more natural acoustic variations.

\textbf{Label characteristics.} Model performance was generally strongest for imprecise consonants and naturalness and weakest for breathiness (Table \ref{tab:spearman_all}).  Imprecise consonants and naturalness had distinct annotation distributions from each other (Fig. \ref{fig:sap_annotation_distributions.png}), suggesting that the difference in model performance is not simply a function of the data distribution.   In addition to the strength of the signal related to the VQD itself, performance could also be impacted by annotator confidence, which might vary with speech category due to length and prosody differences.  Future work could include obtaining annotations from different raters to quantify annotator reliability and assess the impact on results.

\begin{figure}[h!]
  \centering
  \includegraphics[width=\linewidth]{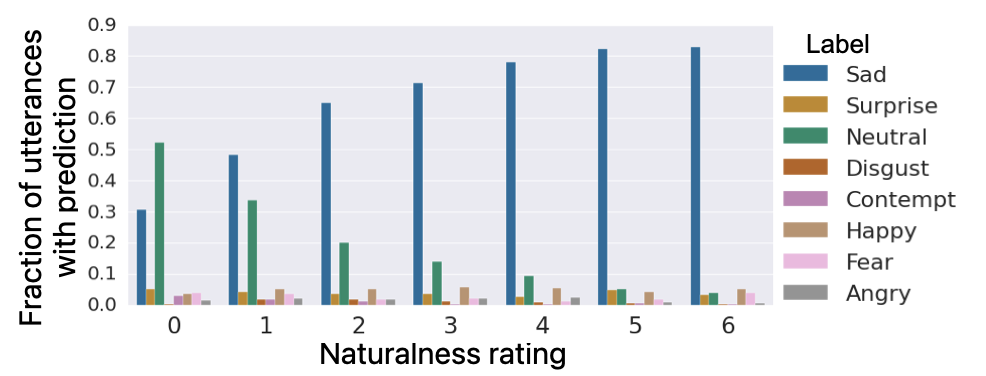}
  \caption{Affective predictions for more atypical speech shows less neutrality and higher sadness} 
  \label{fig:odyssey_sap}
\end{figure}

\textbf{Unseen datasets} 
The probes had strong zero-shot generalization to predicting atypical speech severity on both evaluated datasets (Table \ref{tab:auc_external} and Fig. \ref{fig:external_severity}). 
 Interestingly, breathiness was a stronger differentiator for Italian speech than English speech in classifying severity (Table \ref{fig:external_severity}).  The impact of VQDs on perceived atypical speech severity may vary depending on the properties of a language.  Future work will include evaluation with additional languages including tonal languages.  The CLAP model had the strongest performance on the Italian dataset, suggesting that the general sound-understanding training strategy generalizes particularly well across languages. Despite training on a much smaller dataset, our probes had consistently higher performance than the SpICE models - including for intelligibility (the label learned by SpICE) evaluated on datasets out of domain for all approaches (Tables \ref{tab:spearman_all} and \ref{tab:auc_external}).  Because both our probes and the SpICE probes had low complexity, data and label quality may play a large role.  There were differences in protocols between datasets -- for instance, the data used to train SpICE models was labeled on a coarser per-speaker basis compared to the per-utterance SAP labels. 

\textbf{Affect} Fig. \ref{fig:hubert_ravdess_all} shows zero-shot predictions for each VQD on RAVDESS using HuBERT embeddings, grouped by categorical emotion.  There were clear, understandable trends in the predicted VQDs even though the probes were trained on data that did not have overt emotional content.  For instance, angry speech had lower monoloudness scores, calm speech had less harshness, and sad speech had higher monopitch.  The low intelligibility ratings for calm speech were surprising -- one hypothesis is that this could be due to low volume and less prosodic variations in calm speech.  The relationship between affect and VQD has been well explored, particularly in speech sciences \cite{anikin2020moan, gobl2003role} but less so in machine learning based models.  Zero-shot model transferability, as illustrated in Fig. \ref{fig:hubert_ravdess_all} by applying probes trained on neutral atypical speech to affective speech, suggests utility of VQD-based descriptions of speech in ML generally.  However, because many modeling approaches rely directly on annotated affect labels, ML approaches to affect detection may also be biased for atypical speech.  We evaluated the Odyssey affect model \cite{goncalves2024odyssey} on SAP data and found that affective predictions for more atypical speech shows less neutrality and more sadness, despite no purposeful affective variations in speech elicitation or labeling (Fig. \ref{fig:odyssey_sap}).  While the general bias towards sad labels is not unique to atypical speech \cite{ma2023investigating}, the severity of the bias as a function of atypicality suggests particularly poor performance for people with atypical speech; extended evaluation is an area of future work.

\section{Conclusions} The generalizability and interpretability of the trained VQD models across datasets, languages, and tasks highlights the utility of VQDs for modeling non-semantic speech dimensions.  Additionally, VQD models can also act as explanations or interpretations of decisions made by application focused models, as illustrated by the presented affect exploration.  Limitations of this work include only evaluating frozen embeddings with VQD linear probes, and the limited types of tasks.  Future work should include additional approaches like few-shot learning and adapter training, as well as a deeper investigation of connections between VQDs and specific labels across broad tasks including whether there are dimensions and relationships between dimensions unique to specific labels (e.g., affect, atypical speech detection).  Understanding perceptual ties between voice quality and other labels could improve interpretability, generalization, and robustness across speaking style related tasks.



\bibliographystyle{IEEEtran}
\bibliography{mybib}

\end{document}